\documentclass{MRM}
\usepackage{amsfonts}

\articletype{Research Article -- Submitted to Magnetic Resonance in Medicine}%

\received{xx xx xxxx}
\revised{xx xx xxxx}
\accepted{xx xx xxxx}
\begin{document}

\title{Dual-Domain Multi-Contrast MRI Reconstruction with Synthesis-based Fusion Network}

\author{Junwei Yang\textsuperscript{1}}{\orcid{0000-0002-1415-177X}}

\author{Pietro Li\`o\textsuperscript{1}}{\orcid{0000-0002-0540-5053}}

\authormark{Yang \textsc{et al}}

\address[1]{\orgdiv{Department of Computer Science and Technology}, \orgname{University of Cambridge}, \orgaddress{\state{Cambridge}, \country{United Kingdom}}}


\corres{Junwei Yang, 15 JJ Thomson Ave, Cambridge CB3 0FD, United Kingdom. \email{jy406@cam.ac.uk}}



\abstract[Abstract]{
\section{Purpose} To develop an efficient dual-domain reconstruction framework for multi-contrast MRI, with the focus on minimising cross-contrast misalignment in both the image and the frequency domains to enhance optimisation.
\section{Theory and Methods} Our proposed framework, based on deep learning, facilitates the optimisation for under-sampled target contrast using fully-sampled reference contrast that is quicker to acquire. The method consists of three key steps: 1) Learning to synthesise data resembling the target contrast from the reference contrast; 2) Registering the multi-contrast data to reduce inter-scan motion; and 3) Utilising the registered data for reconstructing the target contrast. These steps involve learning in both domains with regularisation applied to ensure their consistency. We also compare the reconstruction performance with existing deep learning-based methods using a dataset of brain MRI scans.
\section{Results} Extensive experiments demonstrate the superiority of our proposed framework, for up to an 8-fold acceleration rate, compared to state-of-the-art algorithms. Comprehensive analysis and ablation studies further present the effectiveness of the proposed components.
\section{Conclusion} Our dual-domain framework offers a promising approach to multi-contrast MRI reconstruction. It can also be integrated with existing methods to further enhance the reconstruction.
}


\wordcount{XXX}


\maketitle


\section{Introduction}
MRI has become a commonly used imaging modality in radiology, due to its non-invasive nature and the ability of acquiring images of high fidelity and excellent soft tissue contrast. However, its application has a noteworthy limitation, which is the slow acquisition speed. This prolonged acquisition duration is attributed to the fact that the full $k$-space has to be sequentially sampled to encode the spatial-frequency information, while each contrast requires separate scanning. Consequently, the long acquisition duration can lead to patient discomfort and the substantial motion artefacts into the acquired images \cite{krupa2015artifacts}. Thus, the advancement of MRI reconstruction algorithms holds significant potential for substantially enhancing data acquisition efficiency and clinical diagnostic accuracy.

MRI reconstruction is a common task aimed at accelerating image acquisition through the utilisation of sparsely sampled data. It can naturally benefit from an optimisation approach that leverages the coherence between two domains: the frequency domain, known as the $k$-space, and the spatial image domain. Previous studies in MRI reconstruction have predominantly focused on acquiring knowledge to map under-sampled data to their fully-sampled counterparts \cite{fessler2010model, griswold2002generalized, huang2014fast, lustig2007sparse, moratal2008k, pruessmann1999sense, ravishankar2010mr, zhan2015fast}, while deep learning-based methods have also been proposed to further improve the reconstruction \cite{aggarwal2018modl,liu2020highly,qin2018convolutional,schlemper2017deep,seitzer2018adversarial,yang2017dagan}. This task plays an important role in enhancing the quality of MR images given limited measurements, and thus can improve the efficiency and the precision of clinical diagnosis and treatment planning.

Recent advancements in medical image reconstruction, particularly those considering learning in both domains, have incorporated distinct sub-networks designed to leverage interrelated cross-domain information. These sub-networks are usually designed to either be connected in parallel \cite{liu2022dual,jun2021joint, zhang2019dual, zhou2022dual, zhou2023dsformer} or sequentially \cite{eo2018kiki, zhou2020dudornet, zhou2022dudoufnet}. Moreover, considering the separate scanning required for multi-contrast MRI acquisition, attempts have been undertaken to reconstruct images of multiple contrasts. In this context, the exploitation of highly-correlated features between contrasts has been explored to enhance the reconstruction of under-sampled contrast images \cite{dar2020prior,ehrhardt2020multi,kim2018improving,xiang2018ultra,yang2022fast,zhao2022jojonet}. In this task, fully-sampled reference-contrast (reference) images can be leveraged to facilitate the reconstruction of under-sampled target-contrast (target) images, while reconstruction can be significantly improved by exploiting the fully-sampled reference contrast that is quicker to acquire. This multi-contrast learning strategy exhibits promise in accelerating acquisition and enhancing reconstruction quality, particularly for highly sparsely-sampled images.

However, there has been limited exploration of the possibilities associated with the simultaneous integration of dual-domain learning and multi-contrast MRI reconstruction\cite{zhou2020dudornet,feng2021donet,souza2020dual}. In existing methods, the main fusion technique typically involves the straightforward concatenation of data from the two contrasts within each domain, without accounting for potential misalignment issues that may arise during multi-contrast MRI acquisition.

To address the aforementioned limitations, our approach focuses on improving the reconstruction of the target contrast by leveraging information from the data in both domains. Additionally, we aim to reduce cross-contrast inconsistencies caused by inter-scan motion, to facilitate a more effective utilisation of the two contrasts. Through extracting complementary information from the reference contrast, potential improvements in reconstruction performance can be achieved, and the overall imaging experience can be enhanced.

In particular, we propose a multi-stage sequential dual-domain learning framework for multi-contrast MRI reconstruction. In the first stage, the cross-contrast synthesis network is employed to learn to generate the corresponding target image from the reference image. This synthesis task allows for more accurate registration to the target image through a similarity-based loss function in an unsupervised manner. Subsequently, the registration network is utilised to align the synthesised images with the under-sampled images, mitigating inter-scan motion. Finally, the multi-contrast MRI reconstruction network is designed to exploit the measurements from both contrasts, with inputs comprising the synthesised and warped data from the fully-sampled target data and the under-sampled target data. This procedure is applied in both the image domain and the $k$-space, with enforced consistency in outputs across different stages to enhance the exploitation of features learned in the two domains.

For evaluation, our proposed framework is evaluated on a public dataset, in which the reference contrast is chosen as T1-weighted (T1w) due to its shorter acquisition time, while T2-weighted (T2w) and T2 fluid-attenuated inversion recovery (T2-FLAIR) serve individually as the target contrast. Our experimental results demonstrate that the inter-domain consistency constraint enhances the reconstruction when compared to optimisation solely in one domain. Furthermore, the inclusion of an additional reference contrast leads to further improvements compared to conventional multi-contrast fusion strategies. Additionally, we present the effectiveness of our multi-stage strategy, showing that incremental improvements in image quality are achieved through the adaptation of the strategies at each step of the reconstruction process.

\section{Methods}
\label{sec:syn_recon}

\begin{figure*}[t!]
    \centering
    \includegraphics[width=0.7\textwidth]{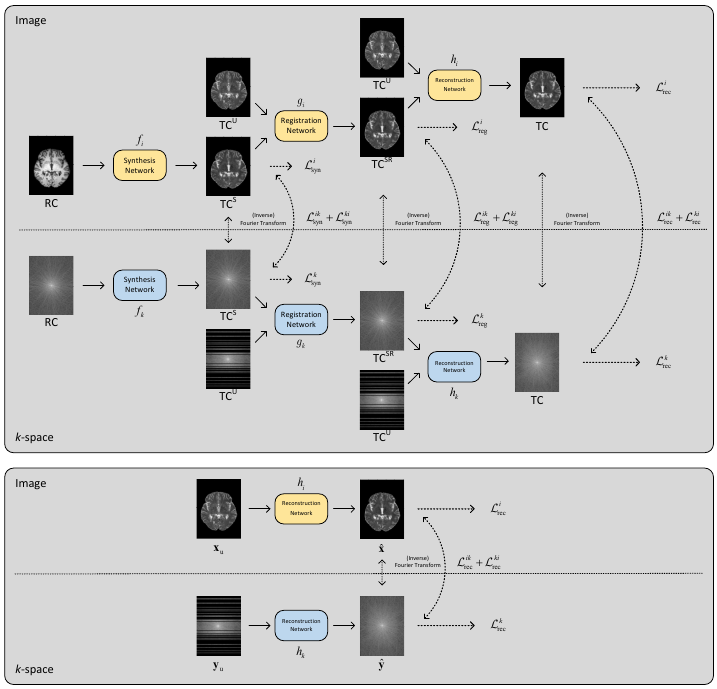} 
    \caption{\ \ The general scheme of our proposed framework for multi-contrast (top) and single-contrast (bottom) MRI reconstruction. The framework includes two symmetrical paths for the image domain and the $k$-space. For multi-contrast MRI reconstruction task, in each domain, given the fully-sampled reference data, a cross-contrast synthesis network is first applied to predict the target image from the corresponding reference data, which are then aligned with the under-sampled target data using a registration network to correct any misalignment induced by inter-scan motion. The aligned data are then concatenated with the under-sampled target data for final reconstruction. The synthesised data and the registered data can be transformed to the other domain by applying the (inverse) Fourier transform to provide extra supervision. For single-contrast reconstruction, only the reconstruction networks are involved, and the consistency between the outputs in the two domains is leveraged to regularise the optimisation. In both cases above, the output from the reconstruction network in the image domain serves as the final output.}
    \label{fig:syn_framework}
\end{figure*}

Our proposed approach is illustrated in Figure \ref{fig:syn_framework}, where three key components are utilised in both domains to enhance the optimisation for multi-contrast MRI reconstruction: the cross-contrast synthesis networks, the registration networks, and the reconstruction networks. At each stage, two networks with identical architectures are employed to learn from data in both domains. By using fully-sampled data as a reference, under-sampled target images can be efficiently reconstructed, thus accelerating the acquisition. Moreover, the framework can also be applied to single-contrast acquisition, as depicted in the lower part of the figure, where only the reconstruction networks are used.

\subsection{Multi-contrast learning}
During the data acquisition process, the fully-sampled $k$-space of a scan with the size of $M \times N$ as $\textbf{y} \in \mathbb{C}^{M \times N}$ can be acquired for single-contrast MRI, while the reference and target scan as $\textbf{y}_{\rm{RC}} \in \mathbb{C}^{M \times N}$ and $\textbf{y}_{\rm{TC}} \in \mathbb{C}^{M \times N}$ can be acquired for multi-contrast MRI, respectively. The corresponding images can be reconstructed from the $k$-space data through inverse Fourier transform $\mathcal{F}^{-1}$, as $\textbf{x} = \mathcal{F}^{-1}(\textbf{y})$, while the reference and target images can be similarly obtained as $\textbf{x}_{\rm{RC}} = \mathcal{F}^{-1}(\textbf{y}_{\rm{RC}})$ and $\textbf{x}_{\rm{TC}} = \mathcal{F}^{-1}(\textbf{y}_{\rm{TC}})$, respectively. A binary under-sampling pattern $\textbf{M} \in \{0, 1\}^{M \times N}$ can be applied to obtain the retrospectively under-sampled $k$-space data as $\textbf{y}_{\text{u}} = \textbf{M} \odot \textbf{y}$ and $\textbf{y}_{\rm{TCu}} = \textbf{M} \odot \textbf{y}_{\rm{TC}}$ for the scenarios of the single-contrast and the target-contrast reconstruction, where $\odot$ represents element-wise multiplication. The corresponding zero-filled reconstructed images can be thus computed as $\textbf{x}_\text{u} = \mathcal{F}^{-1}(\textbf{y}_\text{u})$ and $\textbf{x}_{\rm{TCu}} = \mathcal{F}^{-1}(\textbf{y}_{\rm{TCu}})$. For the single-contrast reconstruction task, the objective is to reconstruct from the under-sampled image $\textbf{x}_\text{u}$ and/or $k$-space data $\textbf{y}_\text{u}$ to obtain the reconstructed high-fidelity image $\hat{\textbf{x}}$. While for the multi-contrast MRI reconstruction task, given the fully-sampled reference images $\textbf{x}_{\rm{RC}}$ and/or $k$-space data $\textbf{y}_{\rm{RC}}$, alongside the under-sampled target images $\textbf{x}_{\rm{TCu}}$ and/or $k$-space data $\textbf{y}_{\rm{TCu}}$, we aim to obtain the reconstructed target image $\hat{\textbf{x}}_{\rm{TC}}$.

In this framework, three types of networks are employed in both domains to facilitate the multi-contrast reconstruction: the cross-contrast synthesis network, the registration network, and the reconstruction network. Depending on the complexity of specific tasks, the architectures can be tailored to accommodate the required capacity.

\subsubsection{Cross-synthesis network}
The cross-contrast synthesis networks are employed in the first stage, specifically designed to learn to synthesise from data of one contrast to the corresponding data of another contrast. The networks are designed to extract information in the reference contrast, which can then be leveraged as complementary information for the under-sampled target data. Additionally, compared to supervised registration using the ground-truth motion between contrasts, the similarity-based comparison can be performed to enable unsupervised learning of the registration networks.

Formally, with the two synthesis networks in the two domains denoted as $f_i$ and $f_k$, these networks can be optimised to produce the synthesised target image and $k$-space data as follows:
\begin{align}
    \hat{\textbf{x}}_{\rm{TCS}} & = f_i(\textbf{x}_{\rm{RC}}), \\
    \hat{\textbf{y}}_{\rm{TCS}} & = f_k(\textbf{y}_{\rm{RC}}). 
\end{align}
Meanwhile, the generated data can also be transformed into the other domain by applying the (inverse) Fourier transforms to obtain the corresponding $k$-space data and the image as:
\begin{align}
    \hat{\textbf{y}}_{\rm{TCSi}} & = \mathcal{F}^{-1}(\hat{\textbf{x}}_{\rm{TCS}}), \\
    \hat{\textbf{x}}_{\rm{TCSk}} & = \mathcal{F}^{-1}(\hat{\textbf{y}}_{\rm{TCS}}),
\end{align}
which can facilitate the enforcement of consistency regularisations.

\subsubsection{Registration network}
With the synthesised target data in both domains, registration networks can be employed to mitigate inter-scan motion. Specifically, these networks are used to align the synthesised data with the under-sampled target data in both domains, denoted as $g_i$ and $g_k$. Since the scans from the evaluation dataset are acquired on the entire brain, the registration networks are designed to learn a rigid transformation, though it can be extended to accommodate non-rigid transformations when necessary. 

Unlike the other two components that were designed to operate separately in the two domains, registration networks are primarily focused on the image domain. In recent years, only a limited number of studies have achieved acceptable performance in learning the registration network using only $k$-space data \cite{kustner2021lapnet,wang2020deepflash}. Representative studies include LAPNet \cite{kustner2021lapnet}, which involves a complex and computationally intensive iterative optimisation procedure. Another approach, DeepFLASH \cite{wang2020deepflash}, focuses on registering with smooth transformations, particularly in low-frequency regions of $k$-space data, though requiring ground-truth transformations for supervision. Consequently, these designs may not be practical in our study and are likely to introduce computational overhead when all three components operate simultaneously.

To enable the registration network for alignment, the registration is devised to warp the synthesised data onto the under-sampled target data. Consequently, the registered data can be acquired in the image domain as follows:
\begin{equation}
    \hat{\textbf{x}}_{\rm{TCSR}} = g_i(\hat{\textbf{x}}_{\rm{TCS}}, \textbf{x}_{\rm{TCu}}).
\end{equation}
On the other hand, for registration in the $k$-space, the registration network with identical architecture is utilised as in the image domain (i.e., $g_i$ is equivalent to $g_k$), and the parameters are shared between the two domains to facilitate information exchange when both branches are considered. Therefore, the registered data can be acquired in the $k$-space domain as follows:
\begin{equation}
    \hat{\textbf{y}}_{\rm{TCSR}} = \mathcal{F}(g_k(\mathcal{F}^{-1}(\hat{\textbf{y}}_{\rm{TCS}}), \mathcal{F}^{-1}(\textbf{y}_{\rm{TCu}}))).
\end{equation}
Similarly, the registered data can be transformed into the other domain, and the transformed $k$-space data and the image can be obtained as:
\begin{align}
    \hat{\textbf{y}}_{\rm{TCSRi}} & = \mathcal{F}(\hat{\textbf{x}}_{\rm{TCSR}}), \\
    \hat{\textbf{x}}_{\rm{TCSRk}} & = \mathcal{F}^{-1}(\hat{\textbf{y}}_{\rm{TCSR}}).
\end{align}

\subsubsection{Reconstruction network}
Finally, the reconstruction networks, denoted as $h_i$ and $h_k$, are employed to reconstruct the images by leveraging information from both contrasts. The alignment strategy employed in the two preceding stages is designed for the reconstruction networks to effectively exploit complementary information from the other contrast. The resulting outputs are subsequently fed into the reconstruction networks as the final stage, and the reconstructed data can be obtained in the two domains as:
\begin{align}
    \hat{\textbf{x}}_{\rm{TC}} & = h_i (\hat{\textbf{x}}_{\rm{TCSR}}, \textbf{x}_{\rm{TCu}}), \\
    \hat{\textbf{y}}_{\rm{TC}} & = h_k (\hat{\textbf{y}}_{\rm{TCSR}}, \textbf{y}_{\rm{TCu}}).
\end{align}
Similarly, the outputs from the two networks can be transformed into the other domain to obtain the $k$-space data and the image as:
\begin{align}
    \hat{\textbf{y}}_{\rm{TCi}} & = \mathcal{F}(\hat{\textbf{x}}_{\rm{TC}}), \\
    \hat{\textbf{x}}_{\rm{TCk}} & = \mathcal{F}^{-1}(\hat{\textbf{y}}_{\rm{TC}}).
\end{align}

\subsection{Dual-domain learning}
Considering the sequential structure of the framework, components can be updated to enhance the task of multi-contrast MRI reconstruction. In our preliminary experiments, we attempted an end-to-end training approach to simultaneously optimise all the networks in both domains. However, this approach proved challenging to converge and demanded powerful hardware resources. As a result, in each stage, we adopt an incremental strategy to optimise the networks progressively, which not only enhances efficiency but also enables better utilisation of outputs from the preceding stage.

To achieve this, as illustrated in Figure \ref{fig:syn_framework}, various loss functions are defined at each stage to facilitate comprehensive bidirectional information exchange between domains. This enables the networks to aggregate and leverage cross-domain information effectively. Consequently, the optimised networks can benefit from the mutual information shared through bidirectional regularisation.

More specifically, in the first stage where the cross-contrast synthesis network is optimised, the objective functions can be defined for optimising based on data in the image domain, $k$-space, and consistency regularisations, as follows:
\begin{align}
    \mathcal{L}_{\rm{syn}}^{i}  & = \|\hat{\textbf{x}}_{\rm{TCS}} - \textbf{x}_{\rm{TC}}\|^2_2, \\
    \mathcal{L}_{\rm{syn}}^{k}  & = \|\hat{\textbf{y}}_{\rm{TCS}} - \textbf{y}_{\rm{TC}}\|^2_2, \\
    \mathcal{L}_{\rm{syn}}^{ik} & = \|\hat{\textbf{x}}_{\rm{TCSk}} - \textbf{x}_{\rm{TC}}\|^2_2, \\
    \mathcal{L}_{\rm{syn}}^{ki} & = \|\hat{\textbf{y}}_{\rm{TCSi}} - \textbf{y}_{\rm{TC}}\|^2_2.
\end{align}
When the dual-domain learning strategy is employed, the loss function can be defined to include all the loss terms above as:
\begin{equation}
    \mathcal{L}_{\rm{syn}} = \mathcal{L}_{\rm{syn}}^{i} + \alpha \mathcal{L}_{\rm{syn}}^{k} + \beta (\mathcal{L}_{\rm{syn}}^{ik} + \alpha \mathcal{L}_{\rm{syn}}^{ki}),
\end{equation}
where the hyper-parameter $\alpha$ is introduced to adjust the weight of loss terms in $k$-space, balancing their different magnitude scales compared to the values in the image domain, while the other hyper-parameter $\beta$ serves as a regularisation term that controls the contribution of the cross-domain consistency constraints on the transformed outputs in the two domains.

In the second stage, the synthesised data is warped to align with the under-sampled target data by minimising its difference with the ground-truth target data. The loss functions utilised in the two domains, along with the regularisation loss terms, are defined as follows:
\begin{align}
    \mathcal{L}_{\rm{reg}}^{i}  & = \|\hat{\textbf{x}}_{\rm{TCSR}} - \textbf{x}_{\rm{TC}}\|^2_2, \\
    \mathcal{L}_{\rm{reg}}^{k}  & = \|\hat{\textbf{y}}_{\rm{TCSR}} - \textbf{y}_{\rm{TC}}\|^2_2, \\
    \mathcal{L}_{\rm{reg}}^{ik} & = \|\hat{\textbf{x}}_{\rm{TCSRk}} - \textbf{x}_{\rm{TC}}\|^2_2, \\
    \mathcal{L}_{\rm{reg}}^{ki} & = \|\hat{\textbf{y}}_{\rm{TCSRi}} - \textbf{y}_{\rm{TC}}\|^2_2.
\end{align}
Therefore, the total loss function and can similarly defined to include all the terms above as:
\begin{equation}
    \mathcal{L}_{\rm{reg}} = \mathcal{L}_{\rm{reg}}^{i} + \alpha \mathcal{L}_{\rm{reg}}^{k} + \beta (\mathcal{L}_{\rm{reg}}^{ik} + \alpha \mathcal{L}_{\rm{reg}}^{ki}).
\end{equation}

Finally, in the third stage, the reconstruction networks are employed in both domains to reconstruct from under-sampled target images with the assistance of extracted complementary information from the reference data. To facilitate dual-domain learning of the reconstruction networks, the loss functions, following the same patterns as in the previous stages, can be defined as follows:
\begin{align}
    \label{chap2:eqn:rec_1}
    \mathcal{L}_{\rm{rec}}^{i}  & = \|\hat{\textbf{x}}_{\rm{TC}} - \textbf{x}_{\rm{TC}}\|^2_2, \\
    \mathcal{L}_{\rm{rec}}^{k}  & = \|\hat{\textbf{y}}_{\rm{TC}} - \textbf{y}_{\rm{TC}}\|^2_2, \\
    \mathcal{L}_{\rm{rec}}^{ik} & = \|\hat{\textbf{x}}_{\rm{TCk}} - \textbf{x}_{\rm{TC}}\|^2_2, \\
    \mathcal{L}_{\rm{rec}}^{ki} & = \|\hat{\textbf{y}}_{\rm{TCi}} - \textbf{y}_{\rm{TC}}\|^2_2. 
\end{align}
With all the loss terms used, the total loss function aggregating the information in the two domains can be defined as:
\begin{equation}
    \label{chap2:eqn:rec_5}
    \mathcal{L}_{\rm{rec}} = \mathcal{L}_{\rm{rec}}^{i} + \alpha \mathcal{L}_{\rm{rec}}^{k} + \beta (\mathcal{L}_{\rm{rec}}^{ik} + \alpha \mathcal{L}_{\rm{rec}}^{ki}).
\end{equation}

With all the components defined for dual-domain learning, the procedures employed in different stages are summarised in Algorithm \ref{alg:al}. Specifically, the synthesis networks $f_i$ and $f_k$ are firstly optimised to generate the corresponding target data from the fully-sampled reference data. Subsequently, with the parameters of the networks $f_i$ and $f_k$ fixed, the registration networks $g_i$ and $g_k$ are optimised based on the inputs partially generated by the synthesis networks. Finally, the parameters of the networks in the first two stages are fixed, and the reconstruction networks are optimised based on the inputs partially generated by the synthesis and registration networks.

It is worth noting that the parameters from preceding steps are intentionally set to be fixed in the second and third stages. While it is possible to simultaneously optimise networks from previous steps, we consider two key factors that argue against such a strategy: 1) Joint optimisation may result in memory overhead; and 2) The varying complexity of different tasks can shift the focus of optimisation and potentially degrade the performance of previously trained networks.

\begin{algorithm}     
    \SetKwInOut{Input}{Input}\SetKwInOut{Require}{Require}\SetKwInOut{Output}{Output}
    \caption{\ Our proposed method}
    \SetAlgoLined
    \Input{Training set $\{\textbf{x}_{\rm{TC}}, \textbf{y}_{\rm{TC}}, \textbf{x}_{\rm{RC}}, \textbf{y}_{\rm{RC}}\}^{N}$; Under-sampled target data $\{\textbf{x}_{\rm{TCu}}, \textbf{y}_{\rm{TCu}}\}^{N}$} 
    \Require{Synthesis networks: $f_{i}$ parameterised by ${\theta_1}$, $f_{k}$ parameterised by ${\theta_2}$; Registration networks: $g_{i}$ parameterised by ${\theta_3}$, $g_{k}$ parameterised by ${\theta_4}$; Reconstruction networks: $h_{i}$ parameterised by ${\theta_3}$, $h_{k}$ parameterised by ${\theta_4}$; Hyper-parameters: $\alpha$, $\beta$} 
    Initialise $f_{i}$, $f_{k}$, $g_{i}$, $g_{k}$, $h_{i}$, and $h_{k}$ \\
    \While {not converged} {
    Optimise ${\theta_1}$ and ${\theta_2}$ based on $\mathcal{L}_{\rm{syn}}$
    }
    Fix ${\theta_1}$ and ${\theta_2}$ \\
    \While {not converged} {
    Optimise ${\theta_3}$ and ${\theta_4}$ based on $\mathcal{L}_{\rm{reg}}$
    }
    Fix ${\theta_1}$, ${\theta_2}$, ${\theta_3}$, and ${\theta_4}$ \\
    \While {not converged} {
    Optimise ${\theta_5}$ and ${\theta_6}$ based on $\mathcal{L}_{\rm{rec}}$
    }
    \Output{Optimised networks $f_{i}$, $g_{i}$, and $h_{i}$}
    \label{alg:al}
\end{algorithm}

\begin{figure*}[t!]
    \centering
    \includegraphics[width=0.9\textwidth]{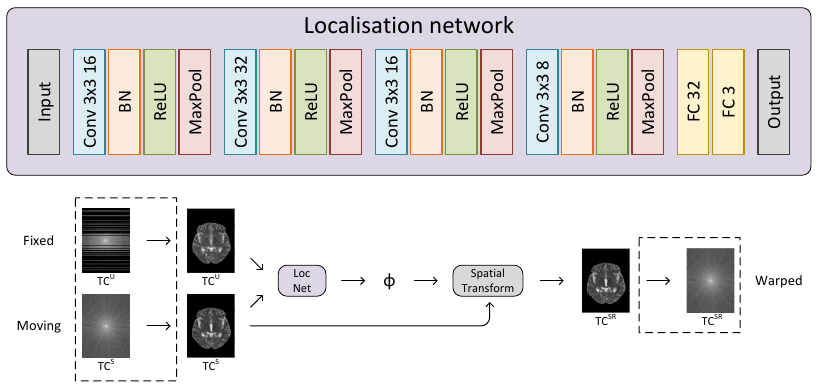} 
    \caption{\ The detailed architecture of the registration network. The localisation network (top) is responsible for estimating the transformation parameters $\phi$ based on the misaligned image pair. Using the estimated motion, the moving image (synthesised target image) is transformed to generate the warped image, which is aligned with the fixed (under-sampled target) image. It is worth noting that, as the registration network is designed for the image domain, the components enclosed by dashed boxes are only applied to the registration network in the $k$-space branch.} 
    \label{fig:reg_framework}
\end{figure*}

\subsection{Experiments}

\subsubsection{Dataset}
To evaluate the proposed framework, we utilised a dataset from the publicly available MICCAI Multi-modal Brain Tumor Segmentation (BraTS) challenge 2019 \cite{menze2014multimodal}. This dataset consists of a large-scale collection of brain MRI data, consisting of 335 preprocessed volumes for each contrast. For each subject, we extracted axial slices for T1w, T2w, and FLAIR contrasts, and cropped them to dimensions of $192 \times 192$ to remove background regions. Since the dataset is processed to ensure alignment across contrasts, to simulate inter-scan motion, we applied random rigid transformations, with random rotations ranging within $\pm10^\circ$ and random translations within $\pm15$mm along both the horizontal and vertical directions.

Following intensity normalisation, which scales the pixel values to a range between 0 and 1 for each slice image, we retrospectively obtained $k$-space data by applying the Fourier transform. A Cartesian Gaussian under-sampling pattern was used to achieve acceleration rates of $4\times$ and $8\times$, with 6 centre lines always acquired to ensure that the important low-frequency information can be obtained. In our experiments, the T1w contrast was selected as the reference, while T2w and FLAIR were chosen as the target contrasts.

In total, the dataset consists of 2010 pairs of randomly selected slices, which were then randomly split at the subject level, with 1407 pairs allocated for training, 201 for validation, and 402 for testing. However, it is important to note that the motion-augmented images were only applied for optimisation in the second and third stages, as the synthesis networks were specifically designed to learn pixel-to-pixel mappings between contrasts.

\subsubsection{Model details}
To demonstrate the effectiveness of our proposed dual-domain constraints and multi-contrast learning strategies, we evaluate our framework using specifically designed network architectures. Inspired by the success of DuDoRNet \cite{zhou2020dudornet}, which employed the same CNN-based architecture for learning to reconstruct in the two domains, we adhere to the same design principle in our framework. However, different architectures are adopted in different stages, due to the varying complexity levels of tasks.

More specifically, the U-Net architecture \cite{ronneberger2015u} is employed as the synthesis network. In the architecture, the inputs consist of two concatenated channels representing the real and imaginary parts of the complex-valued input data, and the output has the same size as the input representing the synthesised data of another contrast.

Regarding the registration networks, we have designed a simple CNN-based architecture to address the complexity of rigid registration. The architecture is depicted in Figure \ref{fig:reg_framework}. In the localisation network, which predicts the transformation parameters, we use the moving image (synthesised target contrast) and the fixed image (under-sampled target contrast) as inputs. This network includes four convolutional layers with the following numbers of feature maps: 16, 32, 16, and 8, while kernel sizes of 3 are employed. Moreover, after each convolutional layer, batch normalisation, ReLU activation, and a 2 $\times$ 2 max-pooling layer with a stride of 2 are applied. Finally, a fully-connected layer with 32 units is used to extract information from the generated feature maps, followed by another fully-connected layer that produces a vector of three elements (representing parameters of translation in two directions and the angle of rotation), as the estimation of cross-contrast motion. These parameters are then used to transform the synthesised images, aligning them with the under-sampled target images, as the output of the network.

In the final reconstruction stage, we also employ the U-Net architecture \cite{ronneberger2015u}. When using multi-contrast MRI data as input, the two contrasts are concatenated along the channel dimension, resulting in a total of four channels. For single-contrast MRI reconstruction, data from only one contrast is used as input. Furthermore, following the output layer that generates the reconstructed target data, we include the data-consistency layer proposed in \cite{schlemper2017deep}. This layer is used to replace incorrect measurements that have already been acquired, thus enhancing data fidelity.

\begin{table*}[t!]
    \centering
    \resizebox{2\columnwidth}{!}{
        \begin{tabular}{lllllllll}
            \hline
                             & \multicolumn{2}{c}{T1w + 1/4 T2w} & \multicolumn{2}{c}{T1w + 1/8 T2w} & \multicolumn{2}{c}{T1w + 1/4 FLAIR} & \multicolumn{2}{c}{T1w + 1/8 FLAIR}  \\ \cline{2-9} 
                             & PSNR           & SSIM             & PSNR           & SSIM             & PSNR           & SSIM             & PSNR           & SSIM           \\ \hline
            Xiang et al.     & 39.99$\pm$4.81 & 0.989$\pm$0.006 & 35.19$\pm$3.87 & 0.974$\pm$0.0011 & 40.95$\pm$3.76 & 0.989$\pm$0.006 & 36.92$\pm$3.80 & 0.975$\pm$0.015 \\
            DuDoRNet         & 41.86$\pm$3.91 & 0.992$\pm$0.0059 & 35.95$\pm$3.38 & 0.975$\pm$0.017 & 41.92$\pm$4.66 & 0.993$\pm$0.007 & 37.23$\pm$3.66 & 0.977$\pm$0.014 \\
            Ours             & \bf{42.10$\pm$4.51} & \bf{0.995$\pm$0.0040} & \bf{37.06$\pm$3.81} & \bf{0.984$\pm$0.0011} & \bf{43.04$\pm$4.29} & \bf{0.994$\pm$0.0050} & \bf{38.60$\pm$4.51} & \bf{0.980$\pm$0.0010} \\ \hline
        \end{tabular}}
    \caption{\ Quantitative results of the dual-domain reconstruction method with different input formats and under-sampling factors of 1/4 ($4\times$ acceleration rate) and 1/8 ($8\times$ acceleration rate) on the dataset.}
    \label{tab:main_res}
\end{table*}

\begin{table*}[t!]
    
    \centering
    \resizebox{1.9\columnwidth}{!}{
    \begin{tabular}{lllllll}
        \hline
            \multirow{2}{*}{Acceleration}                   & \multicolumn{2}{c}{Image}     & \multicolumn{2}{c}{$k$-space}   & \multicolumn{2}{c}{Dual (i)}   \\ \cline{2-7} 
             & PSNR & SSIM & PSNR & SSIM & PSNR & SSIM       \\ \hline
            $4\times$ & 40.60$\pm$3.98 & 0.991$\pm$0.0055 & 36.11$\pm$3.89 & 0.935$\pm$0.040 & 41.86$\pm$4.57 & 0.992$\pm$0.0050  \\
            $8\times$ & 36.41$\pm$3.91 & 0.981$\pm$0.011 & 33.08$\pm$3.59 & 0.905$\pm$0.050 & 36.92$\pm$3.78 & 0.983$\pm$0.011  \\ \hline
    \end{tabular}}
    \caption{\ The quantitative results of the evaluation on learning using different strategies to leverage information from the two domains. The evaluation includes T2w data for acceleration rates of $4\times$ and $8\times$.}
    
    \label{tab:dudo_res}
    
\end{table*}

\subsubsection{Implementation details}
\label{sec.id}
To evaluate the proposed dual-domain multi-contrast MRI reconstruction method, the three networks are sequentially optimised as outlined in Algorithm \ref{alg:al}. In the optimisations of the three types of networks, the Adam optimiser \cite{kingma2014adam} with the learning rate of $2 \times 10^{-4}$ is applied for all networks.

Regarding the selection of hyper-parameter values, we choose $\alpha$ as $10^{-2}$ through a grid search for $\alpha \in [10^{-3}, 10^{-2}, 10^{-1}]$ conducted on the validation set. Meanwhile, the value also reflects a rough magnitude ratio of the data between the two domains, ensuring a well-balanced optimisation progress, where neither branch dominates the training. For the other hyper-parameter, $\beta$, which determines the strength of consistency constraints, we choose the value of $0.7$ based on a grid search for $\beta \in [0.1, 0.3, 0.5, 0.7, 0.9]$ conducted on the validation set. The framework is optimised using an NVIDIA Tesla V100 graphics card.

For the metrics, we report the PSNR and SSIM between the generated images and the ground-truth images. The SSIM is implemented by following the original definition \cite{wang2004image}. Since the dataset provides segmentation masks for all image pairs, the metrics are computed exclusively for the brain region to better reflect performance in clinically significant areas. It is worth noting that ground-truth images without motion simulation are used to evaluate the synthesis networks in the first stage to evaluate the quality of the synthesised data. For the remaining two stages, the same set of motion-augmented data is employed as ground truth to evaluate performance. The results for outputs from the reconstruction networks in both domains are reported; however, only the networks in the image domain are retained after optimisation for inference.

\subsubsection{Evaluation}
In our experiments, we compare our proposed framework to two state-of-the-art methods to demonstrate its superiority, where the comparison is made for acceleration rates of $4\times$ and $8\times$.

The first method is DuDoRNet \cite{zhou2020dudornet}, for which separate experiments are conducted for dual-domain multi-contrast MRI reconstruction. Due to hardware limitations while ensuring a fair comparison, we set the number of recurrent blocks to 2 and perform optimisation for 1000 epochs on the training set until convergence. Additionally, we consider another baseline method \cite{xiang2018ultra}, originally designed for multi-contrast MRI reconstruction. We employ the U-Net as the backbone network for this method, and referred to as Xiang et al.~in the reported results. 

Meanwhile, additional information from the frequency domain (i.e., $k$-space) is introduced to regularise the optimisation of networks in the image domain, while the incorporation of fully-sampled data from another contrast is employed to further improve the reconstruction. Thus, it is important to understand the contributions of the components proposed in the framework.

The ablation study is firstly conducted to examine the impact of dual-domain learning. To achieve this, the two branches are individually trained with data from their respective domains. Specifically, when optimising the networks in the image domain (i.e., $f_i$, $g_i$, and $h_i$), only the terms $\mathcal{L}_{\rm{syn}}^{i}$, $\mathcal{L}_{\rm{reg}}^{i}$, and $\mathcal{L}_{\rm{rec}}^{i}$ are included in the loss functions of $\mathcal{L}_{\rm{syn}}$, $\mathcal{L}_{\rm{reg}}$, and $\mathcal{L}_{\rm{rec}}$, respectively. Similar settings are applied to training in the $k$-space, where only the loss terms directly associated with $k$-space data are considered.

The study further investigates the impact of introducing another contrast by conducting experiments on the task of single-contrast MRI reconstruction. As depicted in the lower part of Figure \ref{fig:syn_framework}, only the reconstruction networks $h_i$ and $h_k$ are involved in this context. To optimise these two networks, similar loss functions as defined in Equations \ref{chap2:eqn:rec_1}--\ref{chap2:eqn:rec_5} can be considered for single-contrast data. While the key difference lies in the input format of the reconstruction networks $h_i$ and $h_k$, where each network exclusively takes the under-sampled data as input (e.g., $\hat{\textbf{x}} = h_i(\textbf{x}_{\rm{u}})$ for image domain reconstruction). To ensure fair comparison with the multi-contrast MRI reconstruction task, the target contrast (i.e., T2w) is selected as the contrast to be reconstructed.

\section{Results}

\begin{figure*}
    \centering
    \includegraphics[width=\textwidth]{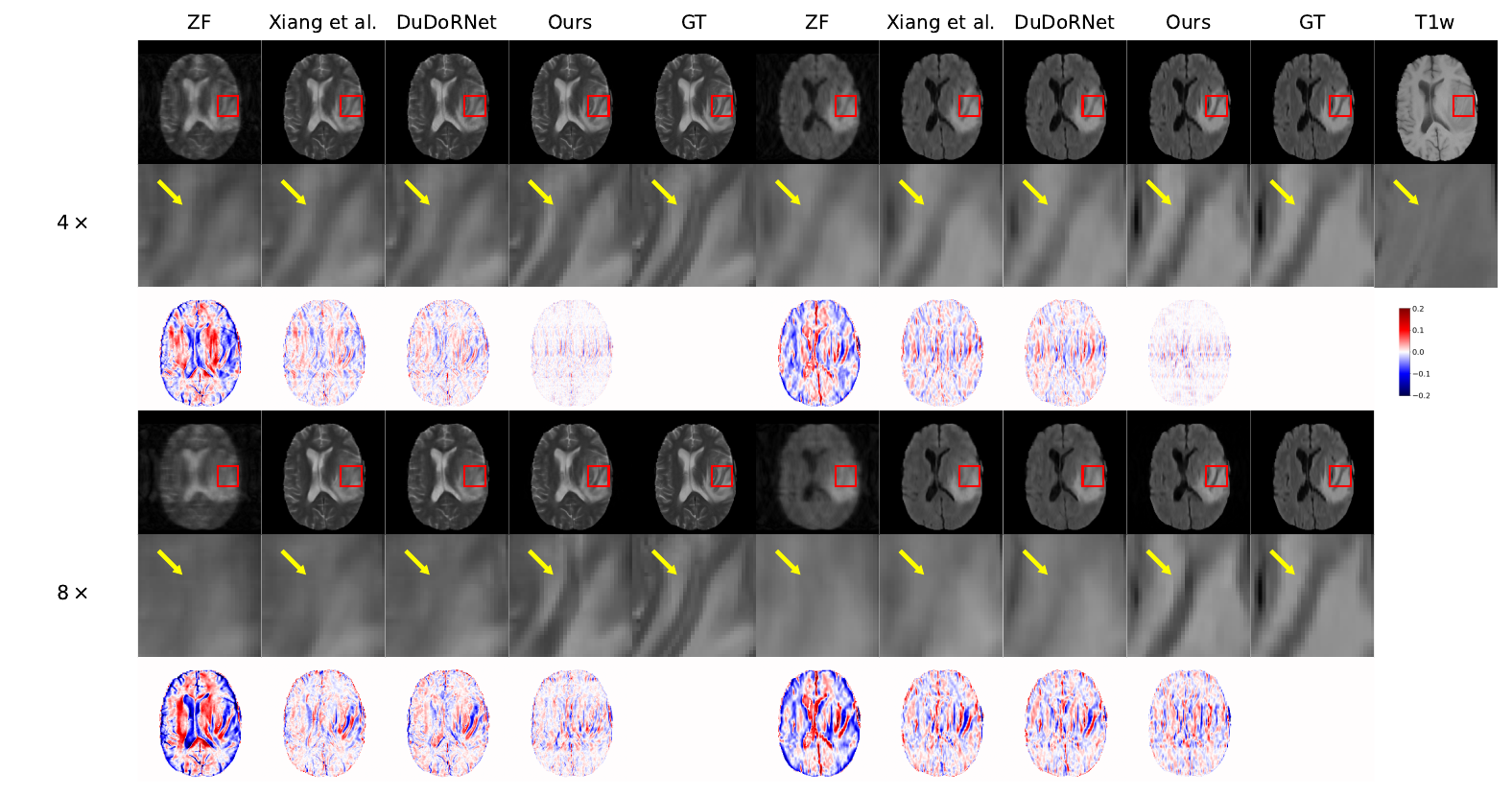} 
    \caption{\ Qualitative results of the comparison made among all considered methods on the dataset under $4\times$ and $8\times$ acceleration rates. ZF represents the under-sampled zero-filled images, and GT represents the ground-truth images.}
    \label{fig:qual_brats}
\end{figure*}

Our framework is designed to optimise in both domains by exploiting the native representation of MRI data, with regularizations that enforce optimisation consistency across the two domains. In this section, we first evaluate the performance of our proposed framework and compare it with the baseline methods. Then, we demonstrate the effectiveness of dual-domain learning by comparing it with single-domain learning methods. Additionally, we present experimental results to highlight the effectiveness of introducing an additional fully-sampled contrast. Next, we demonstrate the influence of our proposed fusion strategy that operates in the two domains. Finally, we investigate the differences in outputs from the two branches to better illustrate our selection of the final output.

\subsection{Dual-domain multi-contrast MRI reconstruction}
The quantitative results of our proposed reconstruction framework are presented in Table~\ref{tab:main_res}. It is evident that our proposed methods outperform state-of-the-art reconstruction algorithms under both acceleration rates for both target contrasts. Moreover, qualitative results are illustrated in Figure~\ref{fig:qual_brats} to visually demonstrate the superiority of our framework. It can be observed that our proposed methods perform generally better for the $4\times$ acceleration rate with overall lower errors. For predictions made with the $8\times$ acceleration rate, although the overall error is similar, it is apparent from the zoomed-in regions that our methods can better recover details.

\begin{figure}[t!]
    \centering
    \includegraphics[width=0.45\textwidth]{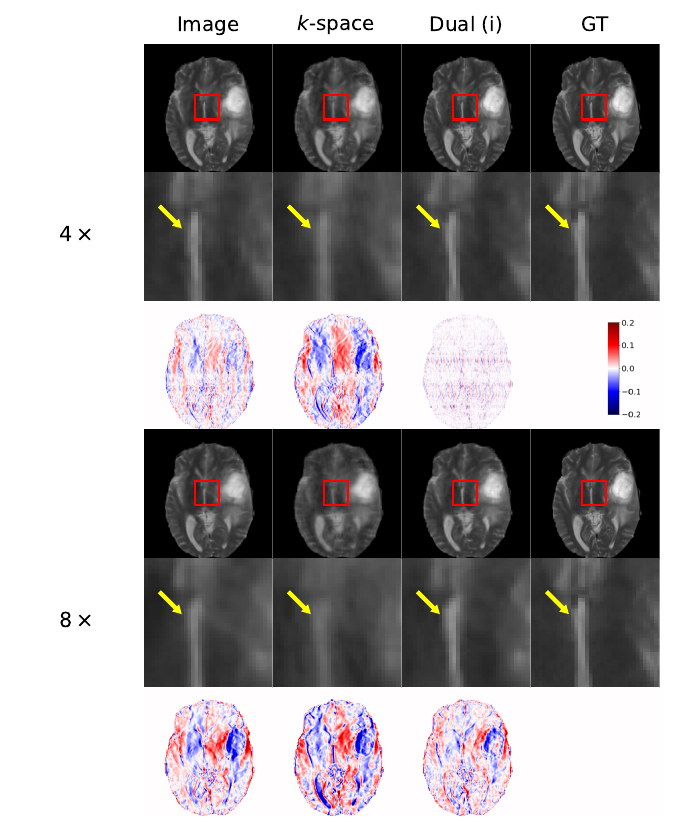} 
    \caption{\ Reconstruction results using data from different domains for our method. The evaluation is conducted on the dataset under both $4\times$ and $8\times$ acceleration rates. Dual (i) represents the output from the image branch when employing the dual-domain training strategy.}
    \label{fig:dudo_res}
\end{figure}

\subsection{Effectiveness of dual-domain learning}
To have a better understanding of the impact of incorporating data from another domain with consistency regularisation, we conducted experiments using a dataset for multi-contrast MRI reconstruction. The reconstruction results are presented in Table~\ref{tab:dudo_res}, with T2w as the target contrast. As indicated, models trained with the proposed dual-domain learning strategy outperform those trained solely in the image domain or $k$-space. Additionally, the Figure~\ref{fig:dudo_res} displays the reconstruction results, highlighting the effectiveness of dual-domain learning for both contrasts and various acceleration rates, as the proposed approach consistently outperforms the conventional learning strategy that focuses on optimising in a single domain. The evaluation conducted on the dataset demonstrates the robustness and effectiveness of the proposed dual-domain learning strategy, for the multi-contrast MRI reconstruction task, across different acceleration rates.

\subsection{Effectiveness of multi-contrast learning}
Furthermore, the impact of leveraging complementary information from another contrast is investigated. Specifically, the evaluation was conducted on the dataset to reconstruct under-sampled T2w data using three different inputs: under-sampled T2w data only, a simple concatenation of fully-sampled T1w and under-sampled T2w data, and the combination of the two contrasts using our proposed fusion strategy.

\begin{figure}[t!]
    \centering
    \includegraphics[width=0.45\textwidth]{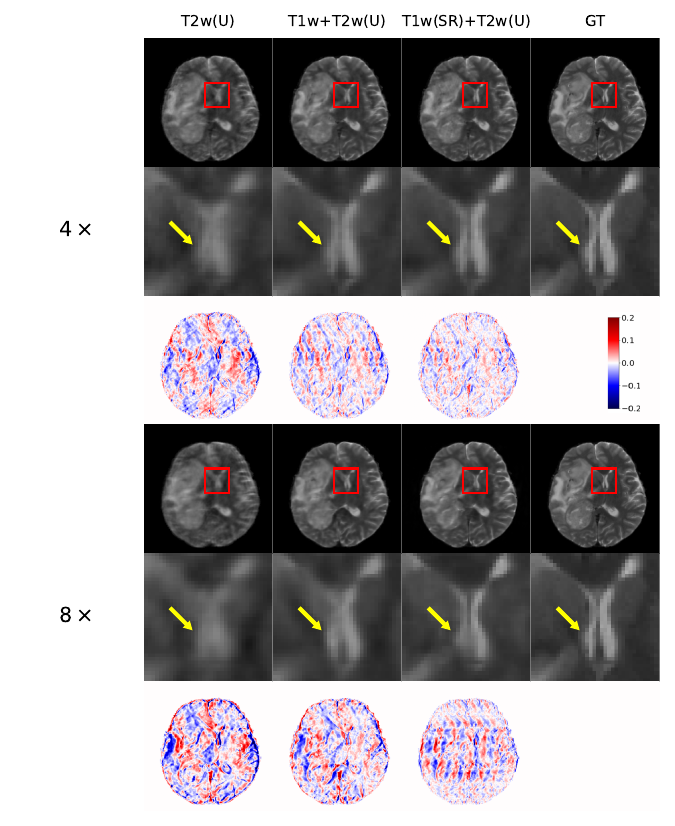} 
    \caption{\ Reconstruction results of using different combinations of data input for the $4\times$ and $8\times$ acceleration rates, including reconstructing from only the under-sampled T2w data, the T1w data combined with under-sampled T2w data, and the two contrasts combined by our fusion algorithms are visualised.}
    \label{fig:mc_res}
\end{figure}

The quantitative results are summarised in Table~\ref{tab:mc_res}. It is evident that introducing another contrast can enhance the reconstruction of under-sampled data, while our fusion strategy, designed to align the reference data with the under-sampled target data optimally, further improves the reconstruction performance. The reconstruction results are also illustrated in Figure~\ref{fig:mc_res}. It is observable that the inclusion of an additional contrast provides noticeable benefits for reconstruction, and our fusion strategy achieves additional performance improvement on top of this.

\begin{table}[t!]

    \centering
    \resizebox{0.9\columnwidth}{!}{
    \begin{tabular}{lllll}
        \hline
            \multirow{2}{*}{Input}                   & \multicolumn{2}{c}{$4\times$}     & \multicolumn{2}{c}{$8\times$}      \\ \cline{2-5} 
              & PSNR & SSIM & PSNR & SSIM \\ \hline
            T2w (U)       & 39.71$\pm$4.94 & 0.988$\pm$0.0063 & 34.97$\pm$3.91 & 0.963$\pm$0.013   \\
            T1w + T2w (U) & 40.20$\pm$4.56 & 0.990$\pm$0.0047 & 36.72$\pm$3.75 & 0.975$\pm$0.011   \\
            T1w (SR) + T2w (U) & 41.86$\pm$4.57 & 0.992$\pm$0.0050 & 36.92$\pm$3.78 & 0.983$\pm$0.011   \\ \hline
    \end{tabular}}
    \caption{\ Reconstruction results for different data input combinations for both the $4\times$ and $8\times$ acceleration rates. These combinations include reconstructing from solely the under-sampled T2w data, the combination of fully-sampled T1w data with under-sampled T2w data, and the concatenation of both contrasts using our proposed fusion algorithms. (U) represents under-sampled, and (SR) represents synthesised and registered.}
    \label{tab:mc_res}
    
\end{table}

\subsection{Effectiveness of sequential learning}

Furthermore, the values of metrics were computed for the intermediate outputs to illustrate the advantages of each stage for both under-sampling rates. The quantitative results for different stages are presented in Table~\ref{tab:seq_res}. Additionally, examples of synthesised intermediate outputs are visualised in the first column of Figure~\ref{fig:dualik}.

As demonstrated in the results, beyond the synthesised images that struggle to capture details in the target contrast, the incorporation of the registration network further aligns the images, enhancing their similarity with the ground-truth target images. This alignment contributes significantly to the reconstruction, compared to the baseline approach without the involvement of the first two stages. It is worth noting that the use of a U-Net architecture for the synthesis network, chosen due to hardware limitations and the consideration of maintaining the consistency of architectures between tasks, may not be capable enough and induce limitations in cross-contrast synthesis. Employing a more advanced backbone network for cross-contrast synthesis is likely to further enhance the reconstruction.

\begin{table*}[t!]
    \centering
    \resizebox{1.9\columnwidth}{!}{
    \begin{tabular}{lllllllll}
        \hline
            \multirow{2}{*}{Stage}                   & \multicolumn{2}{c}{Image}     & \multicolumn{2}{c}{$k$-space}   & \multicolumn{2}{c}{Dual (i)}  & \multicolumn{2}{c}{Dual (k)} \\ \cline{2-9} 
             & PSNR & SSIM & PSNR & SSIM & PSNR & SSIM & PSNR & SSIM      \\ \hline
            Synthesis      & 28.64$\pm$3.45 & 0.955$\pm$0.023 & 26.79$\pm$3.99 & 0.889$\pm$0.061 & 29.03$\pm$3.46 & 0.957$\pm$0.022 & 27.62$\pm$3.82 & 0.943$\pm$0.039  \\
            Registration   & 29.07$\pm$3.25 & 0.956$\pm$0.022 & 27.96$\pm$3.38 & 0.954$\pm$0.023 & 29.24$\pm$3.51 & 0.958 $\pm$0.021 & 27.77$\pm$3.74 & 0.946$\pm$0.035 \\ \hline
            Reconstruction ($4\times$) & 40.60$\pm$3.98 & 0.991$\pm$0.0055 & 36.11$\pm$3.89 & 0.935$\pm$0.040 & 41.86$\pm$4.57 & 0.992$\pm$0.0050 & 36.74$\pm$3.47 & 0.982$\pm$0.032\\ 
            Reconstruction ($8\times$) & 36.41$\pm$3.91 & 0.981$\pm$0.011 & 33.08$\pm$3.59 & 0.905$\pm$0.050 & 36.92$\pm$3.78 & 0.983$\pm$0.011 & 35.12$\pm$3.59 & 0.974$\pm$0.038 \\ 
            \hline
    \end{tabular}}
    \caption{\ Quantitative results of evaluation based on outputs from different stages in our proposed sequential framework. The PSNR and SSIM values are reported for different strategies of exploiting information from the two domains. The evaluation was conducted on the T2w data from the dataset.}
    \label{tab:seq_res}
\end{table*}

\subsection{Selection of outputs for dual-domain learning}
In our proposed dual-domain learning strategy, we simultaneously optimise two branches using the proposed cross-domain constraints to regularise the optimisation. Therefore, the reconstructed output from either domain can serve as the final output of the framework.

Based on the findings from comprehensive experiments conducted in the study of KIKI-net\cite{eo2018kiki}, which investigated the performance of the model optimised in each domain, it was observed that there is a significant performance drop when learning in the $k$-space domain. As a result, we have chosen to include the $k$-space branch solely for training purposes, while inference can be performed in the image domain, also to reduce computational complexity.

\begin{figure}[t!]
    \centering
    \includegraphics[width=0.48\textwidth]{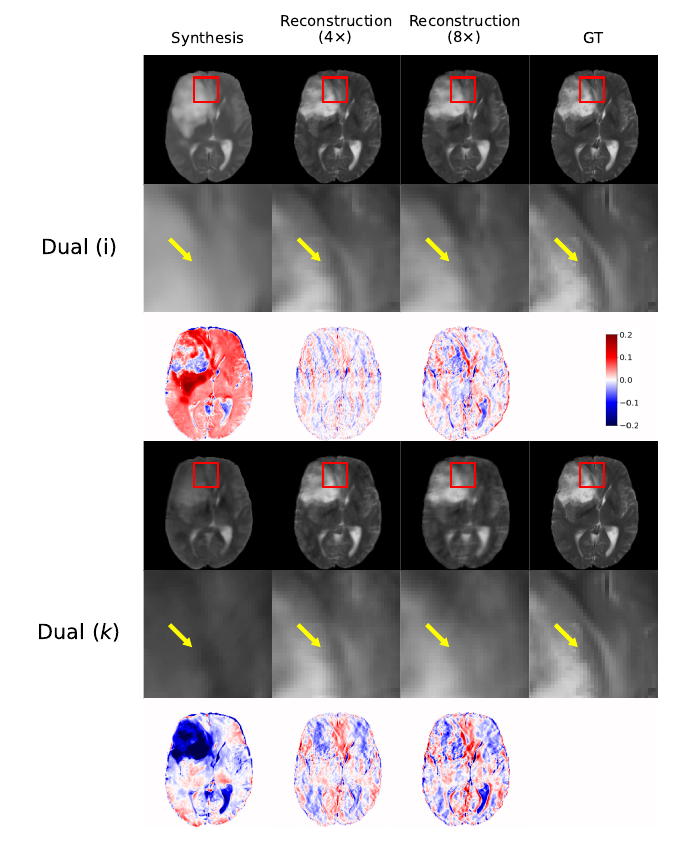} 
    \caption{\ Examples of outputs from different stages and settings for the evaluation on the T2w data from the dataset with the dual-domain learning strategy employed. For each setting, the outputs from both the image (Dual (i)) and the $k$-space (Dual ($k$)) branch are presented.}
    \label{fig:dualik}
\end{figure}

To compare the performance between the two branches in our framework, we conducted a comparison of outputs from different stages. The quantitative results are presented in the last two columns of Table~\ref{tab:seq_res}, and selected examples are illustrated in Figure~\ref{fig:dualik}. It is evident that outputs from the image domain exhibit significantly higher quality than those from the $k$-space branch for all the tasks. Additionally, systematic errors in contrast are captured by the synthesis networks for both domains, addressing the effectiveness of selecting the output from the image domain as the final output.

\section{Discussion}

In this study, we propose a dual-domain framework for multi-contrast MRI reconstruction by exploiting the consistency between the frequency domain, referred to as $k$-space, and the image domain during optimisation. This approach is based on the principle that complementary information from the original acquisition domain can enhance MRI reconstruction that is typically performed only in the image domain. Moreover, considering the potential for information sharing across contrasts, we introduce a sequential pipeline aimed at aligning the fully-sampled reference contrast with the under-sampled contrast. This alignment is designed to enable the model to effectively exploit information from the fully-sampled reference contrast, complementing the under-sampled target contrast, to further improve multi-contrast MRI reconstruction.

More importantly, unlike existing dual-domain learning methods that either employ sequentially-cascaded architectures or individually regularise data from each domain without exploiting on their consistency, our proposed framework incorporates a hierarchical consistency constraint tailored specifically for multi-contrast MRI reconstruction where misalignment can occur in the acquisition of different contrasts.

In our proposed framework, the optimisation in the two domains is regularised at different stages to align the reference contrast with the target contrast. Therefore, the data from both contrasts can be effectively fused for reconstruction. This structured approach stabilises the training process while jointly enhancing reconstruction by leveraging information from both domains.

In our extensive experiments, our proposed framework is compared to state-of-the-art reconstruction methods, for evaluating its performance in the multi-contrast MRI reconstruction task, at acceleration rates of up to $8\times$. Additionally, we have conducted comprehensive ablation studies to evaluate the effectiveness of the individual components designed within our framework. The results demonstrate the superior performance of our proposed method on a public dataset when compared to state-of-the-art reconstruction methods, where the data were retrospectively under-sampled using Cartesian Gaussian random sampling patterns. Furthermore, the introduction of the reference contrast has been shown to be highly effective in enhancing reconstruction when coupled with our proposed multi-stage dual-domain fusion strategy. Notably, the advantages of integrating dual-domain consistency regularisation during optimisation are evident across all the tasks considered at various stages. This highlights the benefits of employing dual-domain regularisation for improving the performance of various MRI-related tasks.

However, in our design, the current general principle is to introduce the $k$-space branch to regularise optimisation in the image domain. Notably, the networks used in the $k$-space branch are not required for inference. This design principle is based on the trade-off between the additional information available in $k$-space and the associated computational complexity. Our experimental results demonstrate the effectiveness of enhancing reconstruction in the image domain through the proposed dual-domain regularisation. Nowadays, learning in $k$-space has been considered as challenging due to highly imbalanced magnitude distribution, and attempts have been made to improve the exploiting information in the frequency domain \cite{han2019k,mildenhall2022nerf}. Therefore, future work may explore the possibility of further integrating the outputs from both branches using specifically designed architectures. This approach could effectively leverage the reconstruction output obtained solely from $k$-space data.

Furthermore, for images acquired using the parallel imaging technique, it is essential to perform coil combination for our proposed framework to function normally. Considering the wide application of parallel imaging in clinical settings \cite{uecker2014espirit,lustig2010spirit} and the potential information loss during the coil combination process, a promising direction for future research is the integration of a deep learning-based module into the existing framework to estimate the coil sensitivity maps for the coil images \cite{sriram2020end,peng2022deepsense}. By leveraging the correlations across coils, this module would provide additional information that can be exploited by the networks at different stages, to eventually improve the reconstruction.

\section{Conclusion}
We have introduced a dual-domain learning framework designed specifically for multi-contrast MRI reconstruction. This framework enables the reconstruction of multi-contrast MR images and $k$-space data, where the fully-sampled reference contrast provides complementary information to enhance the reconstruction of under-sampled target-contrast data. The sequential-structured fusion strategy, which aligns data from different contrasts, enhances the utilisation of the reference contrast. Compared to state-of-the-art algorithms, our proposed framework consistently exhibits superior performance, indicating its potential for more accurate assessments and improved patient comfort.

\subsection*{Conflict of interest}

The authors declare no potential conflict of interests.

\bibliography{MRM-AMA}%
\vfill\pagebreak

\end{document}